# Additive Manufacturing of Nickel Based Superalloys

Apoorv Kulkarni
*Graduate student, Michigan Technological University, Houghton 49931USA*

*Abstract* – **This paper reviews the use of relatively new manufacturing method called as additive manufacturing, most often mentioned as 3D printing in fabrication of high performance superalloys. The overview of the article describes the structure – property – processing – performance relationship of the fabrication process and the superalloys. The manufacturing methods such as Electron Beam Melting, Laser Beam Melting and Direct Energy Deposition used to fabricate commercially available alloys are explained. The microstructure / grain structure resulting from directional building, complex thermal cycles is discussed. An overview of the properties of the superalloys and their performance as well as applications is presented**.

## I. INTRODUCTION

ASTM defines additive manufacturing as layer by layer deposition of materials and joining them, directly from the co-ordinate data obtained from a 3D model. Additive manufacturing is also known by many names such as 3D Printing, Freeform fabrication, Rapid Prototyping etc. [1]. Earlier development of additive manufacturing was focused on manufacturing prototypes from complex polymers. Thus, the name rapid prototyping. Recently there has been a development in manufacturing metal parts through additive manufacturing route. This process facilitates production of complex, custom metal parts which cannot be produced by conventional methods [2] . The advantages of additive manufacturing processes include reduced time for the component manufacturing/ development, little to no waste, flexibility in manufacturing parts with directional solidification, control of the porosity of the part and the ability to produce finer microstructures due to rapid heating and cooling cycles which is usually difficult by the conventional manufacturing methods [3].Most of the metal additive manufacturing technologies use feedstock which is then melted by a concentrated heat source such as laser or electron beam up to a certain depth to fuse it layer by layer and cooled afterwards to build a solid part. The feedstock used may be in the form of wire or powder [4].

Materials used for metal additive manufacturing undergo through very complex thermal processing, i.e. subsequent heating and cooling cycles. The microstructure developed for these components differs from the conventionally manufactured components. Also, there is a difference between the properties of the parts due to the processing. A lot of work is done with alloys such as Ti-6Al-4V. [1] Nickel base superalloy components have been the new focus of development with additive manufacturing processes where the applications have high precision requirement with low volume such as precision engineering, aerospace and medical applications [3].

In this article, the methods used for additive manufacturing of metals/ superalloys, the microstructure of the alloys due to the directionality of the fabrication process, and properties and performance of the superalloys is discussed.

## II. TECHNIQUES FOR ADDITIVE MANUFACTURING OF METALS

Two methods based on the feedstock used for the manufacturing can be classified as Powder Bed Fusion processes (PBF) and direct metal deposition (DMD) or direct energy deposition (DED) [4]. Powder Bed Fusion is among some of the first commercial processes of Additive manufacturing. Powder bed fusion process uses thermal source to fuse the powder particles layer by layer. Different mechanisms are used to spread the powder layer over the build plane [5]. More methods are developed based on the basic Powder Bed Fusion process such as Laser Beam Melting (LBM), Electron Beam Melting (EBM), Direct Metal Laser Sintering (DMLS), Laser Metal Fusion (LMF). These processes utilize similar approach of melting/ sintering the powder layer fusing it with the previous layer using various sources of hear such as Laser, Electron Beam [4].

The Direct Metal Deposition process manufactures parts by melting the metal simultaneously as it is deposited. These methods use a more focused heat source. Unlike the powder bed processes, this process does not melt the material powder that already present on the build platform. The usual mechanism contains two types of nozzles, one houses the metal delivery system which may be powder of wire. The other type of nozzle houses the energy sources which immediately melts the metal that is coming out of the nozzle while following the building path/trajectory fed by the CAD model [5]. Several types of additional processes have been developed with this method such as Laser Metal Deposition (LMD), Direct Energy Deposition (DED), Laser Engineered Net Shaping (LENS)etc. [4].

The two widely used processes used for manufacturing of metal alloy components are the Laser Beam Melting (LBM) or Selective Laser Melting (SLM) and Electron Beam Melting (EBM) [2].

### A. Selective Laser Melting

Laser Beam Melting (LBM) or Selective Laser Melting (SLM) is one of the most promising methods used for Manufacturing metal alloy components. Any additive manufacturing process starts with a 3D CAD model. The model is then sliced into 2D layers of a layer thickness decided by the energy source used for melting the powder [6]. The powder bed is created by rolling the powder on the build platform by using rollers or cassettes. The usual thickness the powder layer is several powder particles. [2] The cross-sectional area is then scanned by rastering the laser spot over the fed 3D CAD model. High power lasers such as CO2, Yttrium are used as the energy source [6]. Then the build plat form is lowered by the same dimension of the layer thickness, another layer of the powder is rolled over the build platform and the process is repeated till the part is manufactured [5]. The average layer thickness for laser melting methods ranges from 20 $\mu$m – 100 $\mu$m [4].

After the metal powder is fed by the hopper or a reservoir, the roller uses a device called recoater to make sure that the powder is spread with uniform thickness [4] [5]. The laser beam is then rastered with scan speed of maximum 15m/s [4]. The scan speed depends on the absorption of the radiation of the energy source into the material as well as the layer thickness that is to be manufactured and other parameters [5]. The thickness of the wall of the structure built depends on the spot size of the laser. The typical spot size ranges from 50 $\mu$m – 180 $\mu$m [4]. The transfer of energy from the laser to the powder layer is affected by factors like absorption, conduction, convection heat transfer, scattering, the flow of fluid when the powder is melted. The major factor is the energy density which controls the layer thickness and the scanning speed. Higher scanning speed translates to lower energy density resulting in unstable melt pool of the powder. Which ultimately delivers poor surface roughness [4].

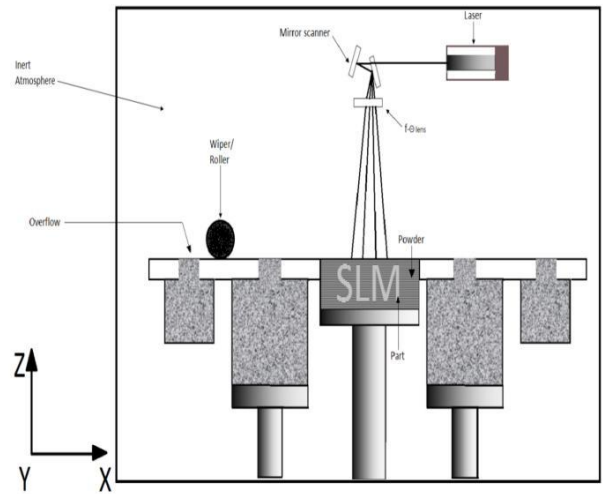

*Figure 1: Schematic of Selective Laser Melting Process [6]*

The primary features of the SLM process are the when concerned with metallurgical point of view is that, when the powder is heated to high temperature gradients, it causes non- equilibrium conditions at the solid/liquid interface. The raster of the laser beam causes cooling of the melt pool which causes it to transform from liquid to solid. Thus, changes in the microstructure of the are observed. Finer microstructures can be formed by increasing the cooling rates sufficiently. The grain structure of the part is controlled by the layer that is solidified before the current layer. The relation of processing parameters for SLM is shown in the Figure 1 [6].

### B. Electron Beam Melting

In Electron Beam Melting (EBM) the powder bed is created by similar method as Laser Beam Melting (LBM). The heat source used is electron beam instead of laser beam which is generated by an electron gun. As described in figure, the electron beam is accelerated with a certain accelerating voltage and then is focused using electromagnetic lenses. Before starting the build, the powder bed is preheated with the help of defocused beam. With the help of electron beam, higher temperatures can be achieved. This helps sinter the powder before staring the actual build scan. The beam current 5mA – 10 mA and scan speed 10mm/s are used for the subsequent scans [4]. The build is often conducted in an inert gas environment like Argon to prevent oxidation.

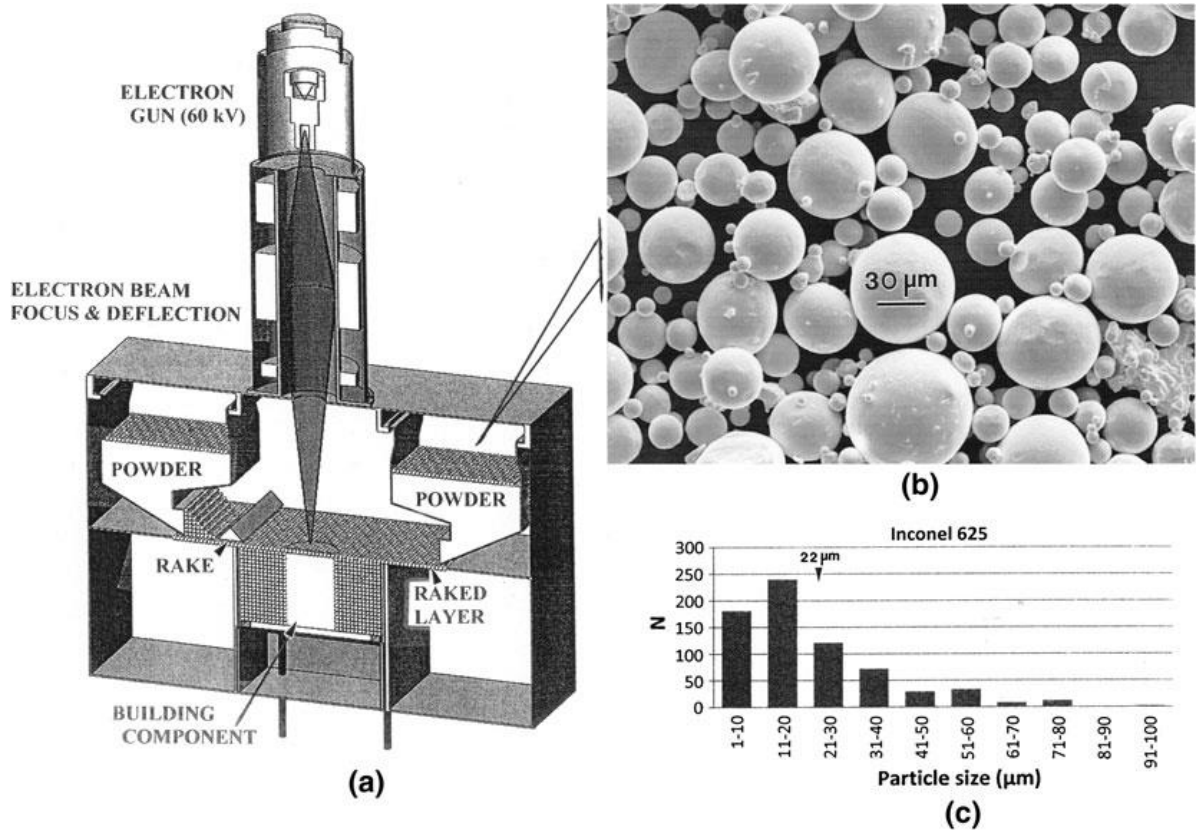

*Figure 2: (a) Schematic of an EBM system (b) SEM image of the precursor powder (c)Histogram of particle sizes of the powder [7]*

### C. Direct Metal Deposition

With Direct Metal Deposition (DMD) or Laser Metal Deposition (LMD) or Direct Energy Deposition (DED) the part is built while simultaneously feeding the metal. The metal can be fed in the form of wire as well as powder. The metal powder/ wire is fed via nozzle, the nozzle can be a single nozzle or a coaxial nozzle surrounding the energy source [5]. LMD is more flexible in terms of the build parameters. Such that the layer thickness can be varied from 40 μm to 1 mm, spot size can be varied from 0.3 mm to 3 mm and the scanning speeds from 150 mm/min to 1.5 m/min. The reason that the system is flexible is that the build plate is stationary, and the deposition head moves. The deposition head is like a 5-axis head which can move around. Some systems have been developed where the head remains stationary and the build plate moves [4]. LMD has wide usage as it can be used to repair cracks, broken parts in turbine systems and gear mechanisms which generally cannot be repaired by conventional methods. LMD can also be used in combination with subtractive methods such as milling machine and a new type of hybrid manufacturing technique has been developed.

## III. FEEDSTOCK FOR ADDITIVE MANUFACTURING OF METALS

Most common materials used for additive manufacturing consist of Aluminum alloys, Titanium alloys, Nickel based superalloys. To generalize, any weldable material can be considered as a suitable material for metal additive manufacturing. Also, while using alloy powders it should observed that some alloys crack under higher solidification rates which are achieved while fabricating, are not appropriate. The crystal structures and mechanical properties of the components produced are different from the ones produced by the conventional manufacturing methods [5].

### A. Powder production

Most of the metal powders are produced by atomization technologies which include water, plasma or gas atomization processes [4].

Powder characteristics such as particle size, morphology of the particle and chemical composition are changed with different powder production methods. The powder properties have considerable impact on the final part that is produced by the AM methods. The fluidity and packing factor of the powder properties have influence on the density on porosity of the part [4].

Water atomization process is the simplest and low-cost atomization process. Liquid metal is atomized by high speed water jets while the metal is falling through the atomization chamber. Particle sizes from ~ 5 μm to 500 μm can be produced by this method due to high cooling rates provided by the water. But this might lead to the particles having an irregular morphology. Which might not be suitable for AM as it reduces the packing factor of the powder. Water atomization also leads to higher oxygen content in the powder particles. This leads to oxide formation of the metals which leads to changes in powder flow behavior and the component composition. This may not be desirable for the materials where oxides have detrimental effects on the mechanical properties of the bulk material [4].

This disadvantage can be overcome by using inert environment with gas atomization. The method works for both reactive and non-reactive materials. The selection of gas influences the "development of microstructure" and in turn the microstructure of the bulk material of the part. The metal melt can be produced by two methods either by vacuum induction melting or electron induction melting where an electrode rod is made from the desired metal and then liquified by passing high intensity current through it. This method mostly results in spherical particles of the metal powder [4].

Plasma atomization can also be used to produce the powder required. In plasma atomization a thing wire is torched with plasma due to which it spherodizes and process spherical particles. The powder particles from plasma atomization are more uniform and fine having sizes of approximately 40 μm [4].

## IV. COMPOSITION OF NICKEL SUPERALLOYS

Superalloys are the types of alloys that are able to retain high strength at elevated temperatures. These have a very complex composition compared to the usual alloys. These also showcase much higher resistance to corrosion and oxidation, and excellent resistance to creep and rupture at elevated temperatures. These are particularly suitable for applications which demand creep resistance at higher temperatures such as aircraft, gas turbines, rocket engines, chemical and petroleum plants as Nickel based superalloys are able to retain their strength when exposed to temperatures above 650°C for long times. There are Nickel based, Nickel-Iron based and Cobalt based superalloys [7]. For additive manufacturing methods mostly Nickel based alloys are used.

These superior properties of Nickel based superalloys are achieved by either solid solution strengthening or precipitation hardening. Most of these alloys are contain 10-20%Cr, up to 8%Al and Ti, 5 to 10% Co and other additions of Mo, Nb, Zr, B, W, Ta, Hf.

### A. General Microstructure of Nickel based superalloy

The major phases that are present in the are as follows;

- **Gamma (γ)** – This the continuous matrix with FCC (face centered cubic) crystal structure. This is called Nickel based austenite. The phase is strengthened by solid solution elements such as Cr, Co, Mo, W, Fe, Ti and Al. Due to slow diffusion of some of the elements the alloy exhibits resistance to high temperature creep [7].
- **Gamma Prime (γ')** – This is the primary strengthening phase which can be precipitated by precipitation hardening treatments. The γ' phase is precipitate in the FCC matric of the form $A_3B$. The "A" is comprised of mostly electronegative elements such as NI, Co and Fe, and "B" is constitutes more electropositive elements such as Al, Ti or Nb. Usually γ' is $Ni_3(Al,Ti)$. This phase exhibits the ordered $L1_2$ crystal structure. The difference in the lattice parameter of γ and γ' phase is about ~0.1% which lets γ' precipitate homogeneously in the matrix. The γ' is ductile, thus it provides strength to the matrix without compromising its fracture toughness. As γ' is long-range ordered, the degree of the order increases with increase in temperature. This results in high strength in temperatures up to 800°C [7].
- **Carbides** – Usually carbon is added from 0.02 – 2% by weight. It combines with reactive elements such as Ta, Ti and Hf to form carbides (TaC, TiC, HfC). As a result of heat treatment, the carbides

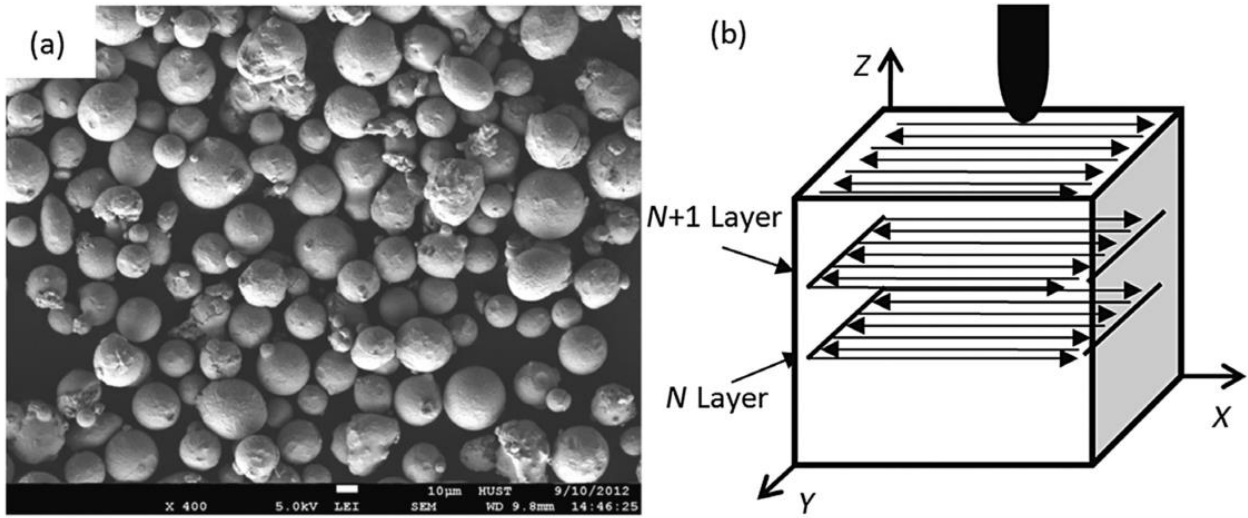

*Figure 3: (a) SEM morphology of Inconel 625 powder (b) illustration of the laser scanning path [9]*

decompose to lower order such as $M_{23}C_6$ and $M_6C$ (M = Metal). The carbides generally have FCC structure. Carbides may be disadvantageous if their amount exceeds a certain percentage, i.e. carbon % is limited to 2%. Below that, carbides are considered to be advantageous when precipitated at grain boundaries, increasing the rupture strength at elevated temperatures. Usually $M_{23}C_6$ form at lower temperature heat treatments (760 - 980°C) and

- **Topologically Close-Packed Phases –** These are usually undesired phases formed during heat treatments. These phases are mostly brittle and form when the composition of the superalloy is not properly controlled. These from in the shape of plates parallel to $\{111\}_\gamma$. These are detrimental because they tend to lower the rupture strengths and ductility.

## V. MICROSTRUCTURE OF ADDITIVELY MANUFACTURED SUPERALLOYS

### A. Macroscopic surface morphology

The generic test pieces of all the researches are found to be cubic blocks with the laser raster pattern as mentioned in the **Error! Reference source not found.**(b). On the top surface a "V" shaped structure was observed similar to welding materials. The adjacent tracks of the raster are found to be overlapping closely with each other. This can be traced to the intense laser heat source which forms a molten pool of the powder as it moves. The powder at the start of the pool keeps continuously melting and the liquid alloy powder quickly solidifies as the laser beam follow the rastering path. This causes difference in the temperature at the head of laser and at the tail of the laser. This difference instigates a contrast in the density and surface tension which generates a convection current in the path of the laser. The convection causes stirring of the molten pool which leads to the formation of the "V" morphology. The angle of the "V" depends on the speed of the lase rastering. The angle gets smaller as the laser speed increases, and in turn the melt pool gets narrower. This summarizes the morphology in the Z direction [8].

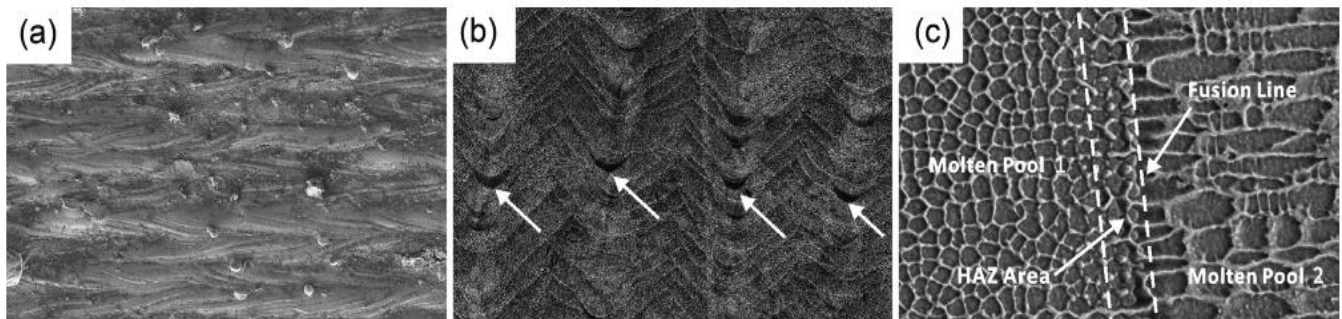

*Figure 4: SEM micrograph of (a) top surface i.e. X-Z direction (b) scales formation in Y-Z direction (c) higher magnification of molten pool and the pool boundary (Inconel 625 via SLM) [9].*

The Y-Z direction shows scale type structure. As the layer thickness of the SLM process is less than 0.05 mm, most of the previous (bottom) layer is re-melted while depositing the subsequent layer. The width of the melt pool is bigger than the spot size of the laser. Similar phenomenon is observed in the Z direction and a slight dip is seen in the middle of the pool. The energy dispersion of the laser into the material exhibit a Gaussian profile. The "scales" seem to match the profile [8].

### B. Dendritic growth

The microstructure of the additively manufactured nickel-based superalloys is highly dependent on the processing parameters and the process history of the part. Hot isostatic processing is performed on most of the parts. As columnar grains formed via additive manufacturing, the microstructure exhibits anisotropy. [9]. The macroscopic properties are substantially affected by the microstructure of the part. The SLM process has very high heat signature which consequences low temperature gradient. As mentioned in [8] "the cooling speed can be as high as $10^6$ K/s for SLM process". Due to this the parts manufactured by SLM exhibit distinct properties than cast superalloys and the wrought superalloys when comparing the microstructure and phase composition [8]. The directional anisotropy that is observed that is due to directional columnar grain growth caused by directional thermal conduction.

Columnar dendrite structure which is the highlight of the additive manufacturing process are formed in the Z direction. Due to the dendrite structure the anisotropy in the microstructure is exhibited. The as processed alloys show a microstructure with significantly elongated grains. Most of the alloys show the dendrite growth in the direction <001> [10]. Most of the superalloys the matrix γ has face centered cubic (FCC) crystal structure. While scanning the laser melts the top of the power layer, the layer on the bottom remains solid (it does not reach the melting temperature). This forms the temperature gradient flowing from top to the bottom. As the heat flux flows into the material, the angle between the direction of the flux and the crystallographic direction <001> is the lowest. Thus, the direction <001> is preferred for the dendrites to grow. Another reason for the directional grain growth can be attributed to the high temperature that is achieved during the SLM/EBM process. The temperatures of the molten pool may reach up to 1800°C. This superheats the melt pool which presents a difficulty in forming homogeneous nuclei. Thus, growth of columnar dendrites is promoted and the dendrites form from bottom to the top. However, temperatures of this order may also lead to high surface tension gradient. This endorses another phenomenon called Marangoni convection. The velocity of the flow of convection an may break the law of columnar dendrite growth and the dendrites may grow in a direction other than <001>. This phenomenon is rare [8]. Researches have shown that the texture of the dendrites is highly influenced by the input laser energy or energy density. Also, the microstructure can be further refined by heat treatment.

Some investigations into phase identification of the constituting phases of the Inconel 718 show that the phases γ and γ' may be the composition of the dendritic structure. Although, distinguishing between the phases is very difficult as found in the XRD characterization, the peaks of γ and γ' overlap [10]. The two phases are difficult to separate in most of the Nickel superalloys as the γ' phase shows ordered structure $L1_2$ and it precipitates coherently with the γ [3].

The solidification in the case of metal additive manufacturing occurs rapidly compared to the conventional methods. This affects he dendrite arm spacing which ranges from ~0.5 µm to 3 µm. Which is 2 orders of magnitude less than the dendrite arm spacing of established casting processes (100 µm – 300 µm). The dendrite spacing is specified by the formula
$$d = a * \varepsilon^{-b}$$
Where d is the dendrite arm spacing; a and b are the material constants. For Nickel based alloys a = 50 µm and b = 1/3. ε is the cooling rate of the melt pool.

MC (Metal Carbides) were formed as precipitates in interdendritic regions. There are segregated during the laser scanning. MCs might help in strengthening and creep resistance if form in right amount, otherwise they are almost always detrimental as they precipitate along grain boundaries and may lead to intergranular fracture. Heat treatment is necessary if MC are precipitate din the microstructure [3].

The amount of the phases such γ, γ' and MCs formed varies with different alloys. The MCs formed during SLM processing of the alloy have less effect on the microstructure. The primary structure of the superalloys was found to be austenitic FCC.

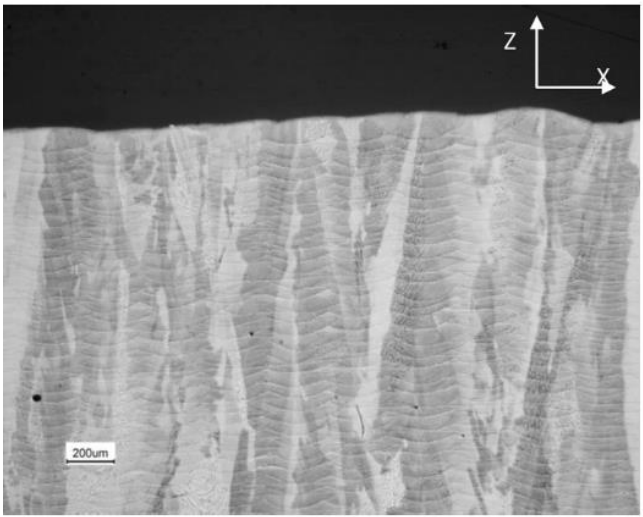

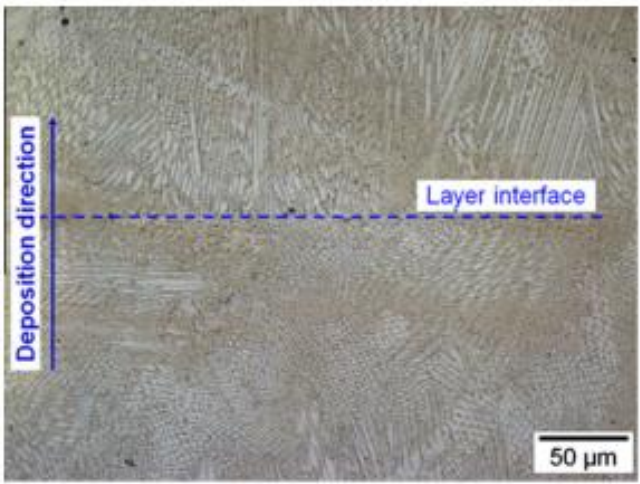

*Figure 5: (a) Dendritic microstructure of Nimonic 263 [11] (b) Densritic microstructure of IN100 [3]*

### C. Effect of Energy Density on the Microstructure

The research [10] suggests that as the energy density (η) of the laser increases the XRD diffraction peaks for the γ and γ' phase broaden, and the intensity of the peaks is significantly decreased. This suggests that as the energy density goes up the crystallite or grain size gets smaller. Also, the diffraction peaks were shifted to a higher angle as η increases. This peak shift indicates lattice distortion, that is increase in the lattice plane distant "d". The lattice distortion is because the γ phase incorporates more uniformly when compared to lower η [10]. Initially looking at the change in the surface morphologies, research [10] considers 4 different cases where the energy density is varied as η = 180 J/m, 275 J/m, 300 J/m and 330 J/m. At the lower values of η the laser rastering tracks were found to be discontinuous. It was also observed that the microstructure had large sized balls with porosities on the surface. As η is increased the laser scan tracks started becoming more continuous and uninterrupted. The balls of the materials reduced in size and the porosities were dispersed and ultimately vanished increasing the density of surface cross section. To compare, when the energy density was increased to 300 J/s, the surface was discovered to be very smooth and the number of metallic globules formed were significantly decreased. When the energy density was further increased to 330 J/m, the density increased to 98.4% as compared to 73.6% at η = 180 J/m. Other factors such as the scan speed, temperature and viscosity of the liquified metal also affect the surface morphology. As the scan speed of the laser is increased, the dwell time decreases and effectively the temperature at the melt pool is decreased. The combined effect of this is higher dynamic viscosity, which in turn increases the porosity of the surface [10].

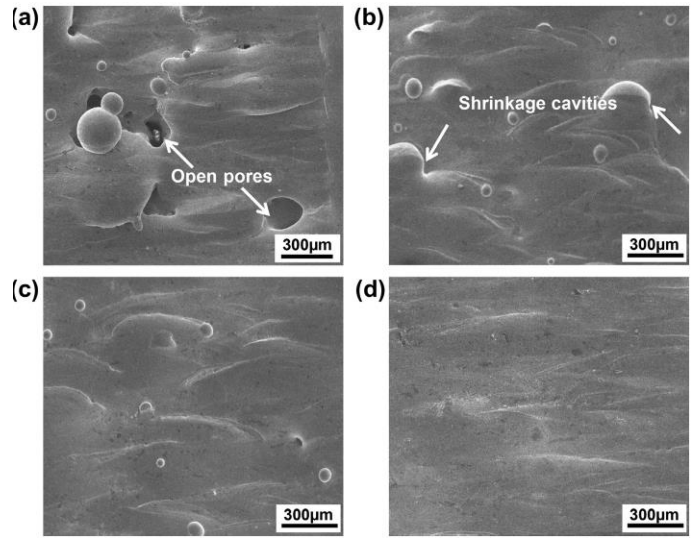

*Figure 6: Surface morphology of Inconel 718 (a) with 180 J/m (b) 275 J/m (c) 300 J/m (d) 330 J/m, (e) Change in density with the change in energy density [10]..*

When looking at the morphology of the dendritic structure, with faster scanning speeds and lower energy density the columnar dendrites formed were coarse and were mostly composed of the γ phase. The columnar grains were said to be "long-cracked in the center of the whole trunk and short-cracked in the interdendritic region". When the scan speed is decreased, and the energy density increased, the dendrites get finer. The dendrites also start to cluster, and it becomes difficult to separate [12].

The energy density not only affects the morphology of the dendritic structure, but also affects the precipitates formed. With high energy density the cooling rates are low. With low cooling rates, more amount of γ' phase is precipitated. With fast cooling rates, less amount γ' phase gets precipitated in γ matrix. In case of little to no precipitation of γ' phase, heat treatment is necessary to segregate the dispersed MCs and to correctly precipitate the γ' strengthening phase [9].

### D. Effect Heat Treatment on the Microstructure

The heat treatment differs for different type of alloys as they have different compositions and different amount of precipitate formed. Heat treatment brings the microstructure of the additively manufactured alloys to the conventional cast alloys. The first stage of any standard heat treatment is high temperature heating. This solutionizes the γ' precipitate phase either partially or totally with the γ matrix in conventional cases [9].

For Nimonic 263, high temperature heat treatment removes dislocations and homogenizes whatever segregations formed. The solutionizing effect might not be seen as compared to the conventional processing of the alloys. As less amount of γ' phase is formed because of rapid cooling during the process, therefore solutionizing the γ' phase becomes redundant. This is then followed by water quenching, to make sure that high number of nuclei are formed. The quenching may again be fast that the γ' precipitates might not form. Annealing helps even distribution of the whatever nuclei are formed. Formation of secondary MC has been seen after annealing of the Nimonic 263. After the heat treatment the width of the dendritic elongated grains is seen to be increased. Thus, the aspect ratio of the columnar grains is decreased [11].

In case of Inconel 718 when it is processed through EBM route, the γ' phase is precipitated in γ matrix. The precipitates align parallel to the building direction. The heat treatment process in this case does not help significantly to refine the microstructure as γ' phase is already precipitated. Aging treatment comes out to be more helpful to get rid of Laves phase [9].

With Inconel 718 manufactured with SLM the microstructure shows dendritic growth and some MCs and Laves phase precipitated in interdendritic region. Strengthening phases such as γ and γ' are not as clearly distinguished in the matrix. Considering this, heat treatment is necessary to help precipitate the strengthening phases properly. The heat treatment also helps with segregation of the MCs and Laves which in turn enhances macrostructural properties [9].

When looked at Inconel 100, the heat treatment pushes the lighter elements in the alloy to interdendritic regions and the heavier elements seem to get congregated in the center of the dendtritic structure. 3 strep heat treatment process is preferred. After the heat treatment 3 types of precipitates primary γ' which are larger in sizes (500 nm dia), secondary γ' precipitates (100 nm) and tertiary γ' precipitates (10 nm). These three types of precipitates are formed at different temperatures during different heat treatments. Primary γ' were formed during solution treatment phase. Secondary γ'phase was formed at cooling temperature of the annealing process. Tertiary γ' phase was formed while aging the additively manufactured part [3].

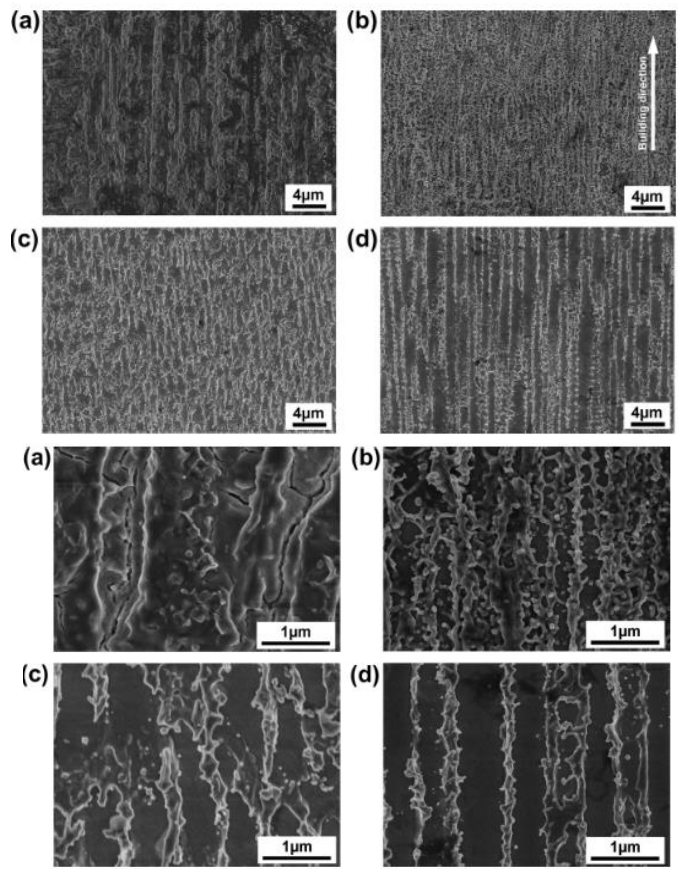

*Figure 7:Dendritic structure formed via AM processes with Inconel 718: (Dentridritic structure gets refined with increase in energy density [11]*

Inconel 625 exhibits somewhat similar results to the heat treatment. The heat treatment makes the grains grow in size, but the grain size seems to be less than the grain size observed with the as cast Inconel 625. After the first heat treatment the grains prefer a certain orientation. The grains finally grow to be rectangular in shape. After the second heat treatment a zigzag type boundary is observed. The zigzag shape of the boundary is important for the ductility.

## VI. PROPERTIES OF ADDITIVELY MANUFACTURED SUPERALLOYS

The properties of the as processed alloys were found to be slightly inferior compared to the as cast alloys. But after heat treatment and aging, the properties improve drastically, and they are on par with the as cast alloys. As we can see from Figure 9 that the Yield Strength (849 MPa vs 738 MPa), % Elongation (22.8 vs 5), hardness (32.5 HRC vs 25 HRC) and Ultimate tensile strength (1126 MPa vs 862 MPa) is greater for as fabricated (SLM) compared to the cast alloy. When again the parts are heat treated the properties show significant increase and are greater than the wrought alloy [6]. From the research we can say that the "superalloy" properties can be retained for the additively manufactured Nickel-based superalloy. The properties of as fabricated alloys show anisotropy suggesting better properties in the build direction (Z-direction) compare to X and Y directions. After heat treatment, the properties seem to approach isotropy.

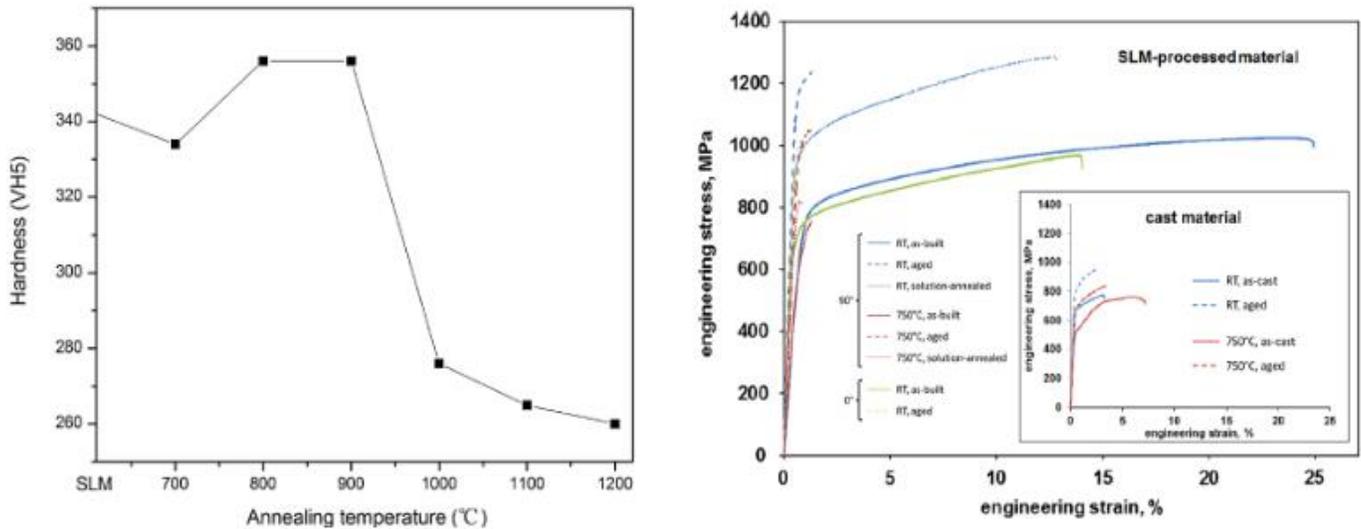

*Figure 8: (a) Hardness of SLM produced Inconel 625 with respect to temperature (b) Stress strain curve comparison of as cast and SLM processed Inconel 939 [6]*

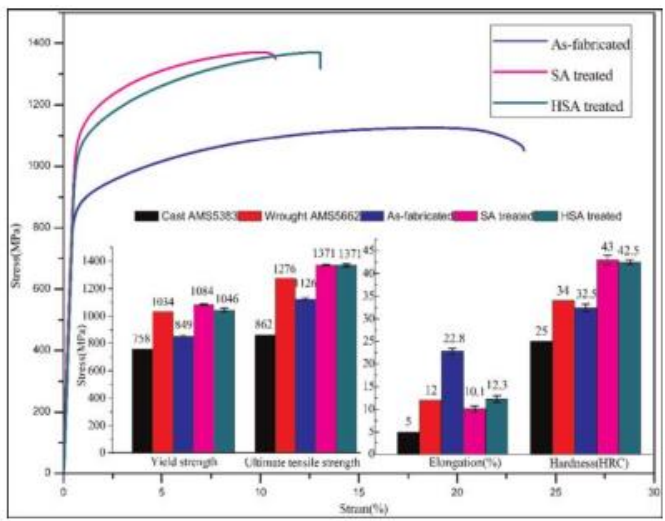

*Figure 9: Stress strain curves comparing the properties of As fabricated, heat treated, cast and wrought Inconel 718 [6]*

## VII. APPLICATIONS OF ADDITIVELY MANUFACTURED SUPERALLOYS

The applications of additive manufacturing are currently restricted to low volume production as mass production of additively manufactured parts is difficult and costly as of now. AM is appropriate for parts having very complex geometry. Usually development and manufacturing of such complex parts is usually more expensive than producing the part with AM. Am processes show promising suture in developing personalized parts and complex structures for aerospace, automobile, energy and biomedical fields. Aerospace components have mostly very complex geometries and their usage is often subject to high strength, high temperatures which asks for high performance materials such as Titanium alloys, Nickel Superalloys, Special Steels etc. The demand for aerospace applications is also less compared to other applications (few thousand parts), therefore AM is very suitable for such applications [13]. Investigations show that the components manufactured via AM posses comparable and sometimes higher strength for the required applications.

Many commercial components use additive manufacturing processes such as LENS, LMD, EBM etc. to manufacture components for satellites, spacecrafts, airplanes, helicopters etc. Complex components such as "turbine blades with thin walls and complex channels" as shown in Figure 10 are manufactured with Inconel 718. Some companies have successfully manufactured and implemented additively manufactured vanes, stators, rotors, geometrically complex parts such as airfoils, blisks, ducts and diffusers. Figure 11 showcases airfoils manufactured from Inconel 738 via LMD process [13].

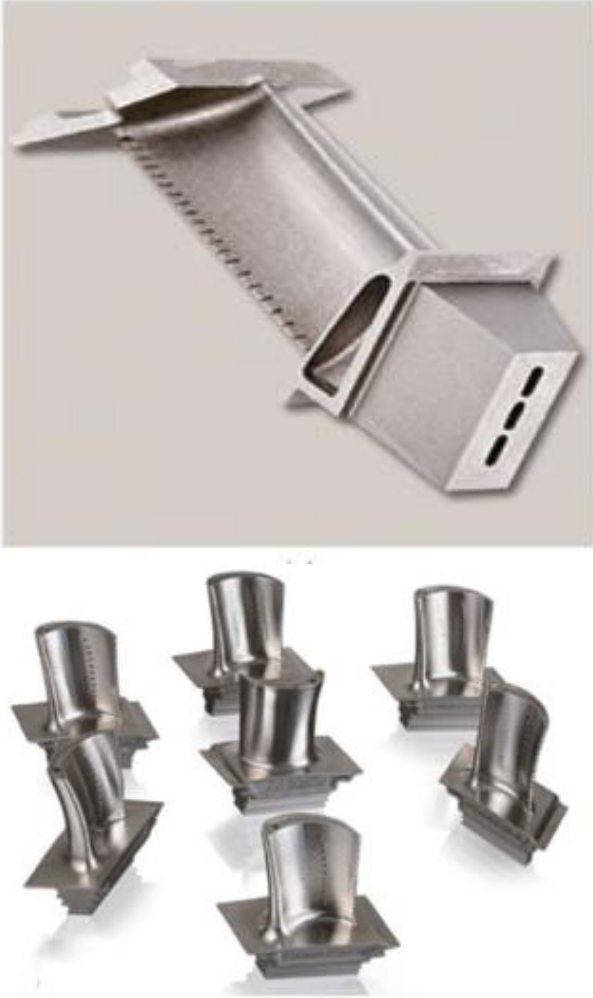

*Figure 10: Turbine blades with thin walls and complex channels with Inconel 718 [14]*

Hybrid manufacturing processes combining "multi axis laser deposition and CNC machining" have been developed which simultaneously deposit and machine the part according to the design.

AM has been applied for repairing of expensive parts. Some components required in aerospace applications are very costly to reproduce if damaged. The damaged component can be repaired by filling the damage via AM processes. Figure 12 shows damaged blisk repaired using LENS [13].

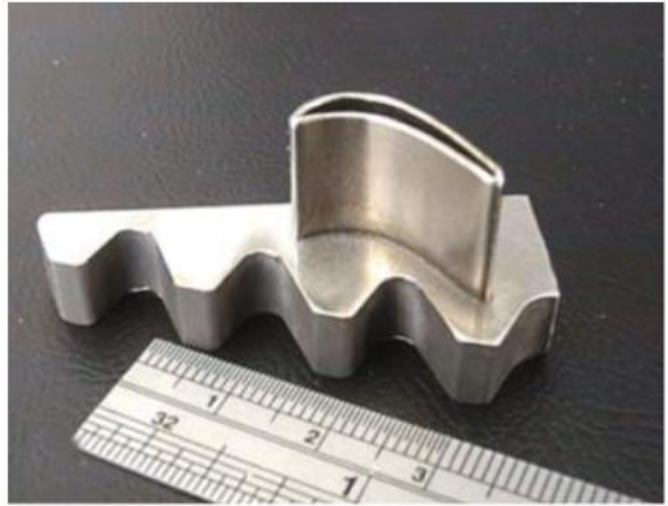

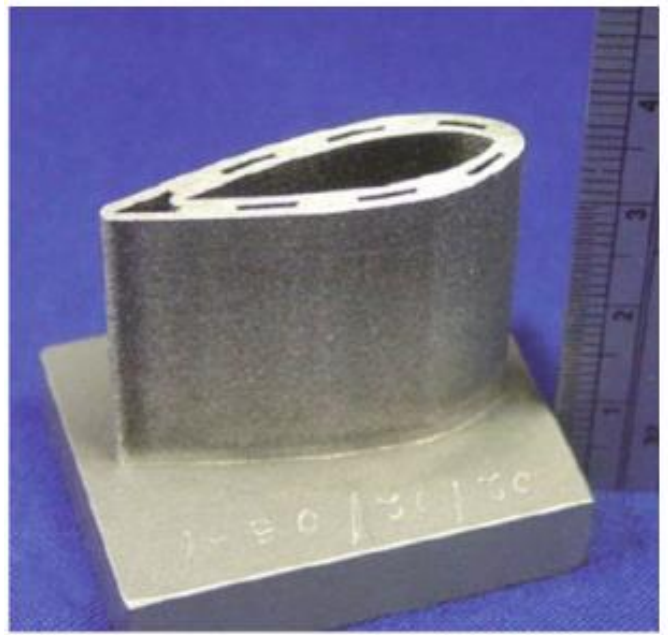

*Figure 11: Airfoils with duct manufactured from Inconel 738 via LMD process [13]*

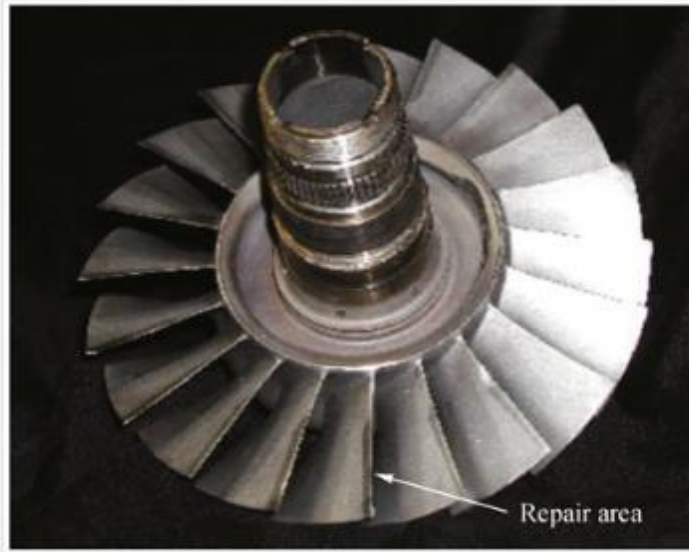

*Figure 12: Damaged area repaired with LENS process*

## VIII. CONCLUSION

To summarize, additive manufacturing is a novel but expensive method to fabricate metal parts. The method is now developed to be able to fabricate complicated and intricate parts that are required for special purpose applications such as aerospace, biomedical, energy and automobile. Nickel superalloys as a material can be a suitable option for theses special purpose applications. The properties of Nickel such as High temperature creep resistance, high temperature rupture strength, retaining the elongation and hardness which makes it viable for this. The microstructure of parts manufactured from Nickel using AM differs from the conventional cast and wrought methods. The microstructure exhibits dendrite like structure (which can also be seen in cast alloys) with aspect ratio of about 5 to 6.5 (much greater than cast). This directional growth of the dendrites affects the isotropy of the properties. Thus, making the part stronger in the build direction of the AM process. The heat treatment done on the parts help increase the width of the dendrites, in turn reducing the aspect ratio and getting the properties closer to isotropy. The cost and time required to develop parts from conceptual stage via AM comes out to be less than using conventional methods. Yet, AM cannot be used for mass manufacturing, but low volume requirement can utilize this method efficiently.